# Photoelectric phenomena in structures based on high-resistivity semiconductor crystals with a thin insulator layer at the semiconductor-metal boundary

P. G. Kasherininov, A. V. Kichaev, and A. A. Tomasov

*A. F, Ioffe Physicotechnical Institute, Russian Academy of Sciences, 194021 St. Petersburg, Russia*

A previously unknown effect—giant spatial redistribution of the electric field strength in a crystal under illumination of the structure — was discovered and investigated in real photoresistors on high-resistivity (semi-insulating) semiconductor CdTe crystals (in metal-thin insulator-semiconductor-thin insulator - metal structures). A new concept is proposed for photoelectric phenomena in high-resistivity semiconductor crystals. The concept is based on the idea that the redistribution of the field under such conditions that the carrier lifetime remains unchanged under illumination plays a determining role in these phenomena. The nature of the effect is described, the dependence of the characteristics of the structures on the parameters of the crystal and the insulator layers is explained by the manifestation of this effect, and ways to produce structures with prescribed photoelectric characteristics for new devices and scientific methods are examined.

A previously unknown effect — giant spatial redistribution of the electric field strength in a crystal under illumination — which in many cases is accompanied by a change in the injection properties of the contacts was discovered and investigated in the structures metal-tunneling insulator-semiconductor-tunnel insulator-metal [M(TI)S(TI)M] based on high-resistivity (semi-insulating) semiconductor crystals. The structures were produced by depositing metal electrodes (M) on the real surface of a crystal (S), coated with a thin layer of the natural oxide — tunnel insulator (TJI). Such structures are widely used as radiation detectors (light, x-ray radiation, -y-rays).[1] The strength of this effect depends on the parameters of the impurity levels in the semiconductor crystal, on the insulator layers of the structure, on the temperature, and on other factors.[2-5]

The new effect is of great interest for both applied and basic research on high-resistivity semiconductors, since the modem theory of photoconductivity in high-resistivity semiconductor crystals was constructed on the basis of investigations of photoelectric phenomena in such structures. The great diversity of unusual photoelectric phenomena in such structures — the noniinearity of the photocurrent $J$ as a function of the illumination intensity $I$ [sub- or superlinear $J = J(I)$], the complicated character of the relaxation of the photocurrents, the residual conductivity (memory of photo-currents), the effect of bias illumination on the magnitude of the photocurrents, the temperature quenching of the photoconductivity, and so on — which, as a rule, accompany one another and are observed in the same samples, is explained in this theory exclusively by the change in the carrier lifetime $\tau(I)$ under illumination, assuming that the spatial distribution of the electric field strength in the crystal remains constant $E(x) = $ const.[6-9] This explanation is rather controversial, it is empirical in nature, and it cannot indicate ways in which structures with prescribed characteristics can be produced.

Based on the effect observed by us we propose a new concept of photoelectric phenomena in such structures. The concept is based on the idea that the photoinduced spatial redistribution of the electric field strength [$E(x)=f(I)$ under the assumption that $\tau(I)=$ const] plays a determining role in these phenomena. This makes it possible to explain from a unified standpoint the diversity of the observed photoelectric phenomena and their interdependence, the dependence of the photoelectric properties of the structure on the parameters of the crystal and insulator layers, and to indicate ways to produce structures with prescribed photoelectric characteristics, which opens up the possibility of practical applications of these phenomena for new devices and scientific methods.

## 1. PHOTOINDUCED SPATIAL REDISTRIBUTION OF THE ELECTRIC FIELD STRENGTH IN THE STRUCTURE M(TI)S(TI)M AND ITS EFFECT ON THE MAGNITUDE OF THE PHOTOCURRENT

It has now been determined that real metal-semiconductor contacts contain a layer of a thin insulator (TT) (natural oxide) which is tunnel-transparent to current carriers (20-50 A thick)[10] so that the high-resistivity structures metal-semiconductor-metai, for which the theory predicts only a small change in the distribution of the electric field upon illumination,[11-13] are in reality structures of the type M(TI)S(TI)M.

Illumination with "characteristic" light $(hv > E_g, V \gg kT/e)$ of one of the electrodes in the symmetric structure M(TI)S(TI)M on a high-resistivity (semi-insulating) semiconductor crystal, in which the Debye length $L_D$ is greater than the thickness of the crystal $d$ $(L_D > d)$ gives rise to the generation of photocarriers near the illuminated electrode $(V$ is the applied voltage, $k$ is Boltzmann's constant, $T$ is the temperature in Kelvins, $e$ is the electron charge, $hv$ is the photon energy, and $E_g$ is the band gap). The generated photocarriers move toward the corresponding electrodes and encounter barriers near the contacts in the form of a thin insulator which they overcome by tunneling.[14]

Stationary through photocurrents are established by means of equalization of the fluxes of photocarriers of each sign which approach the corresponding barrier in the volume of the crystal and pass through it toward the electrode. This process is accompanied by the formation of free photocarriers, which are strictly determined in magnitude and configuration, in the regions near the two electrodes.[19] At the illuminated electrode the increase in the field strength as a result of the accumulation of photocarriers of the corresponding sign occurs in a layer of the crystal with a thickness comparable to $1/\alpha$ ($\alpha = 10^4$ cm$^{-1}$ is the absorption coefficient for light). Elsewhere in the crystal the change in the distribution $E = E(x)$ is due to the charge with the same sign as the illuminated electrode.

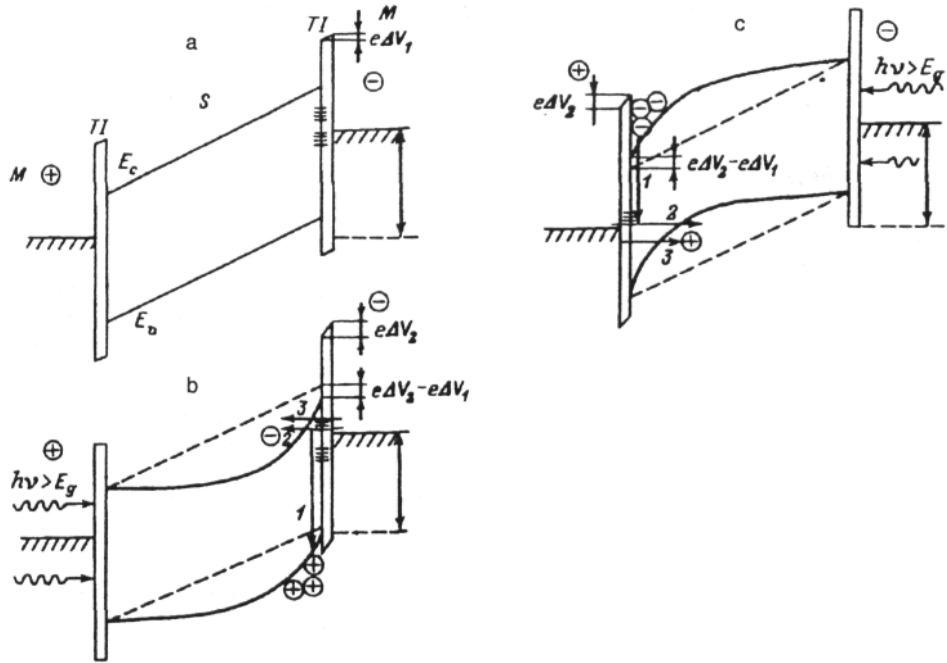

FIG. 1. Energy diagram of a M(TI)S(TI)M structure based on high-resistivity CdTe in the dark (a) and under illumination with the characteristic light from the positive- (b) and negative- (c) electrode sides. M — Metal, S — semiconductor, TI — thin insulator.

When impurity levels are present in the crystal, the free photocarriers which have accumulated near the TI layers can be partially trapped on them. When the concentration of such levels is high, the charge of the trapped carriers can be much higher than the accumulated free-carrier charge, which accounts for the more effective photoinduced redistribution of the field in structures based on such crystals.[4,5]

The difference in the effectiveness of the redistribution of the field for M(TI)S(TI)M structures on different crystals also causes their photoelectric characteristics to differ substantially. A mechanism by which the redistribution of the field influences the magnitude of the photocurrent flowing through the structure is the change in the magnitude of the charge collected on the electrode from each of the photopro-duced electron-hole pairs[20] as compared to the charge with a uniform distribution of the field in the structure.[3]

Another mechanism involves a possible change in the injection properties of the contacts accompanying an increase in the electric field strength in the near-electrode regions of the structure and results in an increase of the current in the external circuit accompanying a redistribution of the field.[4,5] This mechanism is shown in Fig. 1, where the band diagram of the structure M(TI)S(TI)M in the dark (Fig. la) and under illumination with characteristic light from the side of different electrodes (Figs. 1b and 1c) is illustrated. This diagram was constructed on the basis of the experimentally obtained distributions $E = E(x)$ in the structures M(TI)S-(TI)M on high-resistivity CdTe crystals.[2] It is important that when the structure is illuminated, the height of the metal-semiconductor barrier decreases as a result of an increase in the voltage drop in the insulator layer by the amount $(e\Delta V_2 — e\Delta V_l)$; this can give rise to above-barrier carrier injection into the crystal.[19] In addition, the width of the potential barrier at the nonilluminated electrode decreases (Figs. 1b and 1c); this makes possible carrier injection from the side of this electrode by tunneling emission from the surface states into the conduction band (valence band) of the semiconductor through a triangular barrier in the semiconductor (Figs. 1b and 1c, process 2) or tunneling transmission of carriers from the metal through the insulator-semiconductor double barrier (Figs. 1b and 1c, process 3).[21] The field distribution in the crystal influences the magnitude of the photocurrent via all of these mechanisms. As the illumination intensity, the time, the temperature, and other parameters change, one or another mechanism can predominate. This is what determines the characteristic features of the photoelectric properties of these structures.

## 2. FORMULATION OF THE PROBLEM

In the present paper we explain the diversity of photoelectric phenomena which are observed in high-resistivity structures M(TI)S(TI)M with different effectiveness of the photoinduced redistribution of the electric field in them as a result of the formation of spatial electric charges of different magnitudes and configurations. The basic logical stages of the investigation are as follows.

1. For each type of structure the following parameters are measured simultaneously for different intensities of illumination: the magnitude of the stationary photocurrent, the shape of the photocurrent relaxation curve with the lightswitched on (off), and the spatial distribution, $E = E(x)$, between the electrodes of the structure in the stationary state and at different times after the light is switched on (off).

2. The stationary current versus illumination curves and the form of the photocurrent relaxation in each type of structure (assuming there is no carrier injection from the electrodes) are calculated according to the measured distributions, $E = E(x)$.

3. The contribution of injection processes to the current characteristics of each type of structure is estimated by comparing the experimentally measured photoelectric characteristics with the characteristics computed according to Sec. 2.

4. The effect of external illuminations on the photoconductivity of the structures M(TI)S(TI)M via the spatial redistribution of the electric field under the action of these illuminations is studied.

5. The photoelectric phenomena determined exclusively by carrier injection from the electrodes in the presence of photoinduced redistribution of the field in the crystal (super-linear current-versus-illumination curves, residual conductivity, and so on) are studied.

### 3. EXPERIMENTAL PROCEDURE

*3.1. Object of the investigations.* The experimental samples consisted of M(TI)S(TI)M structures which were produced on semi-insulating electro-optic p-CdTe crystals (p=$10^8$ Ω cm) with the concentration of impurity levels varied over a wide range:

a) compensated crystals with the concentration of acceptor and donor type impurity levels W>$10^{15}$ cm$^{-3}$, Γ=300 K;

b) crystals with deep impurity levels (compensated crystals at $T$= 77 K).

The p-CdTe(Cl) crystals were grown by the method of horizontal directed crystallization in the [110] direction with compensation by chlorine ions;[1] the concentration of the impurity levels was changed by introducing deviations from the stoichiometric composition by means of enrichment with cadmium vacancies. These vacancies and their complexes form acceptor and donor type impurity levels.[22] The samples consisted of 0.7x0.7X0.3-cm$^3$ plane-parallel plates, cut perpendicular to the axis of growth; the M(TI)S(TI)M structures were produced by spraying or chemical deposition of optically transparent gold electrodes on an etched 0.7X0.7-cm$^2$ surface which were coated with a thin oxide layer whose thickness could be varied by special oxidation of the surface.[23]

A constant voltage $U$= 100-500 V was applied to the electrodes of the structure. The illumination with a constant light flux or square pulses of the characteristic light (A. = 0.63 or 0.8 μm) was performed through optically transparent electrodes. The results of the experimental studies of structures on pure crystals are presented in part one of this paper.

*3.2. Method for measuring the spatial and temporal distribution of the electric field in the volume of the crystals under an external perturbation.* In the present work we employed a modification of the polarization-optical method that makes it possible to determine with a spatial resolution of 50 μm and a temporal resolution of less than 1 μs the spatial distribution $E = E(x,t)$ in structures based on electro-optic crystals with arbitrary crystallographic orientation and unknown values of the electro-optic coefficients under both stationary conditions and at different times after the external perturbation is switched on (off). The measurements were performed by probing a sample, which was inserted between crossed polarizers, with a narrow light beam that passed through the crystal perpendicular to the direction of the field (beam diameter < 100 μm, λ= 1.3 μm). The beam was directed onto the lateral face of the sample and moved from one electrode to the other with a prescribed step. The intensity of the light flux at the exit after the analyzer was measured with a germanium photodiode.

The time dependences of the output intensity $I^*(x,t)$ of the probe light with the absorbed illumination switched on (off) were measured for each position of the probe beam. The spatial distribution of the Pockels effect between the electrodes of the structure at any time $t$ was reconstructed from a set of these curves.

For such an optical system the output intensity of the beam propagating over a distance $x$ from one of the electrodes is related to the intensity of the electric field $E(x,t)$ by[24] relation (1):

$$I^*(x,t) = I_0^* \sin^2[kE(x,t)], \qquad (1)$$

where $I^*_0$ is the intensity of the incident light, and $k$ is a dimensional electro-optic coefficient that depends on the orientation and the dimensions of the sample.

The space-time distribution $E(x,t)$ was calculated in absolute values from the measured distributions $I^*(x,t)$ using the experimentally established fact that the redistribution of the field in the experimental structures is not accompanied by a significant change in the voltage drop across the semiconductor crystal; i.e.,

$$\int_0^d E(x)dx = V. \qquad (2)$$

The dimensional coefficient $k$ can be determined from Eqs. (1) and (2) and the intensity of the field can be calculated from the formula

$$E(x,t) = \frac{V \sin^{-1}[I^*(x,t)/I_0^*]^{1/2}}{\int_0^d \sin^{-1}[I^*(x,t)/I_0^*]^{1/2}dx}. \qquad (3)$$

*3.3. Method for calculating the current characteristics of the structures.* A method is now available for calculating the magnitude of the charge stored on the electrodes of the structure from any electron-hole pair produced by light at a known point of the crystal for any prescribed distribution of the electric field.[20,3] The magnitude of the photocurrent can therefore be calculated if the illumination intensity, the photocarrier generation function, and the distribution function $E(x,t)$ are known. The computed and experimentally measured photocurrents can differ from one another by the amount of the current component determined by carrier injection from the electrodes of the structure.

The photocurrent density is calculated in the drift approximation under the following assumptions:

1) The structure is illuminated by characteristic light with intensity $I$ from the side of one of the electrodes and the photons are absorbed directly near the illuminated electrode ($x=0$);

2) the parameters $\mu$, and $\tau$ of the semiconductor crystal (electron and hole mobility and lifetimes, respectively) are known and do not change upon illumination.

The charge $Q$ collected on the electrodes of the structure as the induced charge $q$ traverses a distance from $x_1$ to $x_2$ is described, according to Ref. 20 , by the expression

$$dQ = \frac{q(x)}{d}dx, \quad Q(x_1 x_2) = \frac{1}{d}\int_{x_1}^{x_2} q(x)dx. \qquad (4)$$

When the structure is illuminated from the side of the positive electrode, charge is transported through the crystal by holes. The magnitude of this charge will decrease with time according to the relation

where $r$ is the carrier (hole) lifetime, and $t$ is the time elapsed from

$$q(x) = q_0 \exp(-t/\tau), \qquad (5)$$

the moment the charge is generated.

The time $t$ and the coordinate $x$ of the charge carriers in the crystal are related by the relation

$$dt = \frac{dx}{v_{dr}(x)} = \frac{dx}{\mu E(x)}, \qquad (6)$$

where $v_{dr}$ is the drift velocity of the carriers (holes), and $t$ is the hole mobility.

From expressions (4)-(6) we obtain a formula for calculating the charge collected on the electrodes from the charge $q_0$ produced by one photon at the point $x = 0$ (anode)

$$\frac{Q}{q_0} = \frac{1}{d}\int_0^d \exp\left[-\frac{1}{\tau\mu}\int_0^x \frac{dx'}{E(x')}\right]dx. \qquad (7)$$

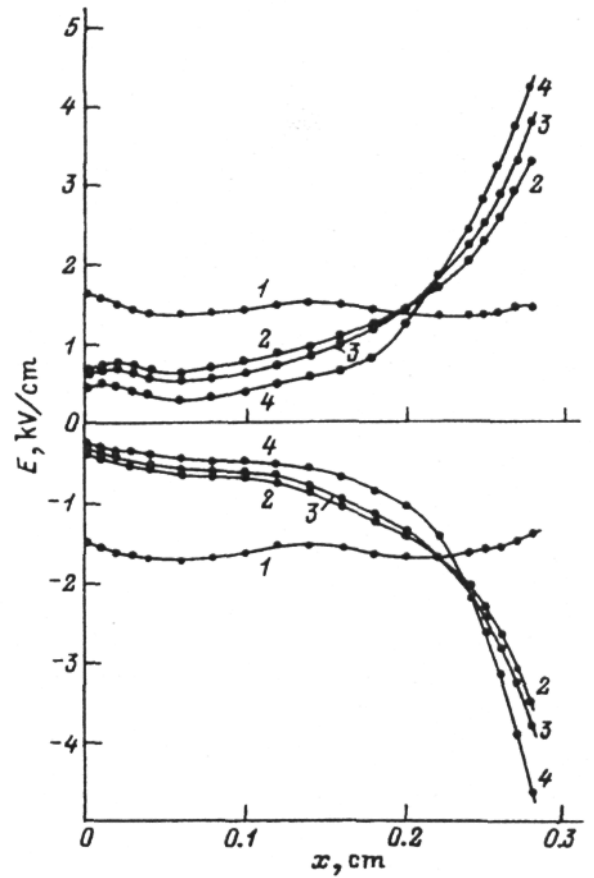

FIG. 2. Stationary distribution of the electric field strength E in the volume of the crystal of a symmetric M(TI)S(TI)M structure with thin insulator layers of the natural oxide under illumination from the positive-electrode (top curves) and negative-electrode (bottom curves) sides. $I$, mW/cm$^2$: 1 — 0; 2 —6; 3— 15; 4 — 95; V=400 V.

For each fixed value of the illumination intensity the photocurrent density has the form

$$J = e\beta(1-R)\frac{Q}{q_0}K_D\frac{I}{h\nu}, \qquad (8)$$

where $e$ is the electron charge, $\beta$ is the internal quantum yield, $R$ is the coefficient of reflection and nonphotoactive absorption of light, and the coefficient $K_D$ takes into account the diffusion of carriers in the direction opposite to the field , and the recombination at the surface.[25]

According to Eqs. (7) and (8), the dependence of the photocurrent density on the illumination intensity, $J=J(I)$, is therefore described in terms of the known distribution of the electric field intensity in the crystal $E(x)=f(I)$.

## 4. PHOTOELECTRIC PHENOMENA IN M(TI)S(TI)M STRUCTURES, BASED ON HIGH-RESISTIVITY PURE CdTe CRYSTALS

*4.1. Distribution of the electric field in a crystal under illumination.* We investigated the structures M(TI)S(TI)M based on pure p-CdTe crystals with electron and hole mobility $\mu_n=800$ cm$^2$/(V-s) and $\mu_p=80$ cm$^2$/(V-s), and the carrier lifetime $\tau=5\text{X}10^{-7}$ s.

The method described above was used to investigate the distribution $E = E(x)$ at different times after the light was switched on/(off), and the action of the operating conditions (the applied voltage, the illumination intensity, and so on) and the parameters of the structure (thickness of the insulator layers) on the distribution of the field strength and the nature of the electric charge giving rise to the field redistribution were studied.

The stationary distributions of the electric field in a symmetric M(TI)S(TI)M structure, with TI layers formed by the natural oxide, with illumination from the side of the positive and negative electrodes are shown in Fig. 2. In the absence "of illumination the electric field is uniform over the thickness of the crystal. Under the action of the light, as the illumination intensity increases, the field strength decreases at the illuminated electrode and increases at the dark electrode, and in structures with insulator layers of the same thickness the distribution $E=E(x)$, which arises under illumination from the side of different electrodes, is sufficiently symmetric.

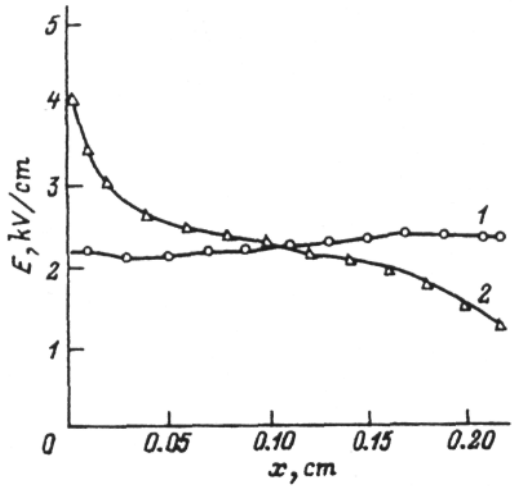

FIG. 3. Distribution of the electric field in the dark in M(TI)S(TI)M structures with TI layers of the same thickness 20-50 A (1) and with a thicker TI layer at the anode (x = 0) (2). V=500 V.

The character of the distribution $E = E(x)$ is determined by charges, of both polarities, of the photocarriers accumulated near the insulator layers on the two electrodes. The magnitude of these charges changes as the thickness of the TI layers changes. The effect of the permittivity of the insulator layers in the M(TI)S(TI)M structure on the electric field distribution in the dark and under illumination was studied by means of special chemical oxidation of one face of the crystal prior to deposition of an optically transparent metallic electrode on it with the thickness of the other TI layer remaining constant. Figure 3 shows the distribution $E = E(x)$ in the absence of illumination in the unsymmetric structure M(TI)S(TI)M with a thicker TI layer at the anode. We see that the electric field is nonuniform over the thickness of the crystal, and that the field is stronger near the electrode with the thicker insulator (near the anode), i.e., decomposition of the impurity levels is observed as the thickness of the TI layer in the crystal increases. As follows from the calculations performed on the basis of the results of this experiment, acceptor-type impurity levels with concentration $\text{Л}^\text{И}\text{О}^{11}$ cm$^{-3}$ are decompensated.

When such a structure is illuminated from the anode side (x = 0), the stationary distribution $E=E(x)$ in the crystal is the same as in a structure with thin TI layers (Fig. 2). The rates at which a stationary distribution of the field is established in these structures, when the illumination is switched on, are also the same (Fig. 4), but the restoration time of the dark distribution of the field, after illumination is switched off, is much longer (by an order of magnitude) in the unsymmetric structure than in the symmetric structure with thin TI layers.

The nature of the electric charge giving rise to the photo-induced change in the spatial distribution $E(x)$ in such structures was determined from the character of the relaxation of the field strength when the illumination is switched on (off). As one can see from Figs. 4a and 4b, in the symmetric structure M(TI)S(TI)M a stationary distribution of the field is established in the semiconductor crystal over a time comparable to the transit time of the carriers (holes) $t<50$ μs; this indicates that the photoinduced charge is apparently determined by the free carriers and (or) recharged shallow impurity levels.

*4.2. Action of external illumination on the magnitude of the photocurrent in the structure.* We studied the effect of external illumination, which changes the spatial distribution $E = E(x)$, on the magnitude of the photocurrent in a M(TT)S-(TI)M structure. The photocurrent was produced by illuminating the structure from the cathode side with a weak modulated flux of the characteristic light ($\lambda = 0.8$ μm) which does not produce any changes in the distribution of the field. The structure was illuminated from the side of different electrodes by a constant light flux of different intensities. In the course of the experiment the spatial distribution $E = E(x)$ in the crystal and the photocurrent as a function of the modulated light flux were recorded.

The photocurrent $J_m(I)$ from the modulated light was calculated, in accordance with Eqs. (7) and (8), as a function of the intensity of the constant illumination on the basis of the measured distributions of the electric field $E(x)=f(I)$

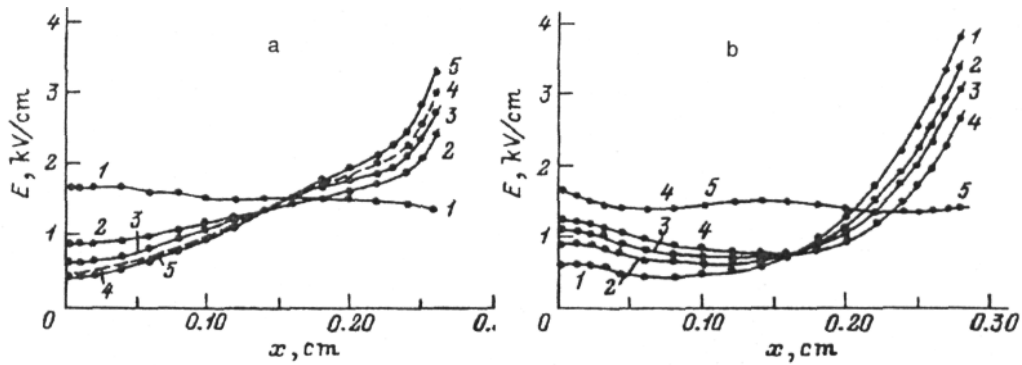

FIG. 4. Distribution of the electric field in the volume of a crystal of a symmetric M(TI)S(TI)M structure at different times ($t$) after illumination is switched on (off) from the positive electrode side, $V$ = 400 V. a — After illumination $I$ = 50 mW/cm$^2$ is switched on, $d$= 0.26 cm; $t$, μs; 1 — 0; 2 — 2; 3 — 5; 4 — 10; 5 — 50. b — After illumination $I$ = 50 mW/cm$^2$ is switched off; $d$= 0.28 cm; $t$, μs, 1 — 0; 2 — 5; 3 — 10; 4 — 20; 5 — 80.

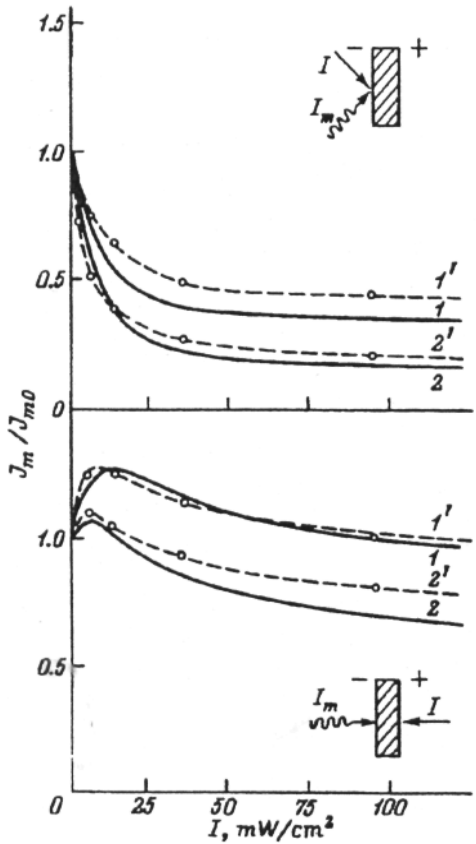

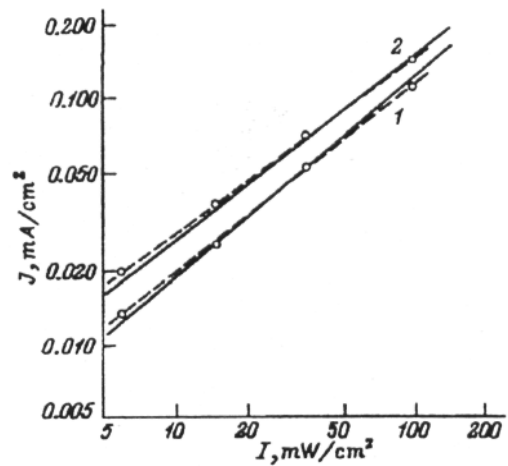

FIG. 5. Photoresponse from a modulated light flux $J_m /J_{m0}$ as a function of the intensity of illumination of a symmetric M(TI)S(TI)M structure from the side of the positive (top curves) and negative (bottom curves) electrodes. V, V: 1,1' — 300; 2,2' — 400. 1, 2 — experiment; 1',2' — calculation based on the measured distribution of the field strength in the crystal. The intensity of the modulated light flux on the anode side is $I_m = 50$ ^W/cm².

FIG. 6. Stationary current-versus-illumination characteristics of a M(TI)S-(TI)M structure: Experimental (solid lines) and computed (dashed lines) from the measured spatial distributions of the electric field strength in the crystal. $V, V:$ 1 — 300, 2 — 400; the structure is illuminated from the anode side.

(Fig. 3) and compared with the corresponding experimental curve. It should be noted that under external illumination and modulated light from the side of one electrode (Fig. 5a) the calculation was performed for $K_D$= const for all illumination intensities, because the change in $K_D$ in this case can be associated only with the change in $E(x = 0)$[25], which is virtually independent of the intensity as a result of the accumulation of carriers of the corresponding sign in a narrow region at the illuminated electrode.

The computed and experimental functions $J_m(I)$ obtained with illumination from the side of the negative and positive electrodes are shown in Fig. 5. The curves agree quite well with one another. In the case where the structure is illuminated from the side of the electrode which is illuminated with the registered (modulated) light, the photocurrent from this flux decreases monotonically as the intensity of the external illumination increases. When the structure is illuminated from the side of the opposite electrode, the recorded photocurrent can increase as the illumination intensity increases.

*4.3. Current-versus-illumination curves.* The stationary current-versus-illumination curve of the symmetric structure M(TI)S(TI)M with thinTI layers, $J = J(I)$, measured experimentally and calculated on the basis of the measured distributions $E = E(x)$ with different intensities of illumination from the positive electrode side (Fig. 2a), is shown in Fig. 6.

The experimental and computed current-versus-illumination curves, as one can see from Fig. 6, agree quite well and are described by the expression $J = AI^n$, where n = 0.7—0.8; i.e., they are sublinear. The agreement indicates that the nonlinearity (sublinearity) of the current-versus-illumination characteristic of this structure is due to the photoinduced spatial redistribution of the electric field in the crystal and the fact that in the entire experimental range of illumination intensities the change in $E = E(x)$ in structures on such crystals is not accompanied by carrier injection from the electrodes.

The contribution of injection currents to the current-versus-illumination characteristics of the structures and the conditions and mechanisms of their appearance have not been studied adequately and are of great interest. In structures on non-electro-optic crystals, where it is difficult to measure the field distribution, the current-versus-illumination characteristic determined by the photoinduced redistribution of the field (in the absence of injection) can be calculated on the basis of measurements of the photore-sponse of the structure from the probe light flux as a function of the intensity of constant illumination $[J_m=J_m(I)]$ from the side of the same electrode (Fig. 5a). The magnitude of this photoresponse is completely determined, as shown above, by the distribution $E = E(x)$ in the crystal and does not respond to injection processes in the structure.

The computed current-versus-illumination characteristic $J(I)$ of the structure is described by the relation

$$J(I) = e\beta(1-R)\frac{Q_0}{q_0}\frac{J_m(I)}{J_m(0)}\frac{I}{h\nu}, \qquad (9)$$

where $J_m(0)$ is the magnitude of the photoresponse from the modulated light with $I = 0$, and $Q_0/q_0$ is the relative magnitude of the collected charge with a uniform distribution of the electric field in the crystal ($I = 0$).[26]

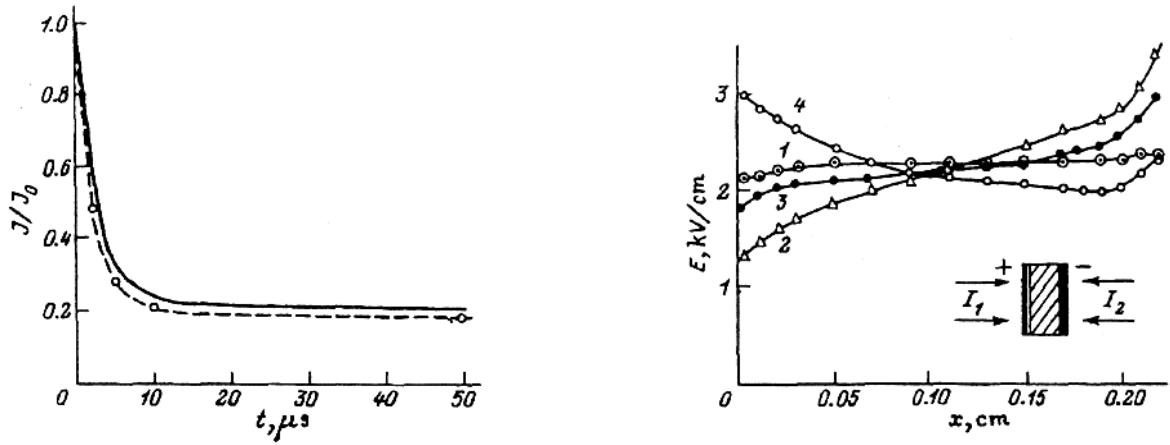

FIG. 7. Photocurrent as a function of time with the structure illuminated by square light pulse from the positive-electrode side. $I$=50 mW/cm$^2$, V=400 V. $J_0$ — Initial photocurrent density under conditions of an unde-formed field. Solid line — Experiment; dashed line — calculation according to known field distributions at different times after the illumination is switched on.

FIG. 8. Distribution of the electric field in a crystal with carrier injection from the dark-electrode side. V=500 V. The intensity of the illumination of the structure from the anode side is $I_1$ = 35 mW/cm$^2$. The intensity of illumination from the cathode side is $I_2$ (mW/cm$^2$): 2 — 0; 3 — 5; 4 — 50. 1 — Dark distribution of the field in the crystal, $I_1$ =0, $I_2$ = 0.

*4.4. Shape of the photocurrent relaxation curve.* We studied the nature of the relaxation of the photocurrent in a symmetric structure M(TI)S(TI)M illuminated from the anode side with a light flux of constant intensity [$I(t)$ = const], which give rise to a spatial redistribution of the electric field $E(x)$=$f(t)$. The distribution of the field in the crystal at different times after the illumination is switched on (on (Fig. 4) was measured, and the form of the relaxation of the photo-current $J(t)$ in the external circuit was calculated from these curves. The magnitude of this current at each point of the crystal is determined by the sum of the photoconduction current $J_p(t)$ and the displacement current $J_0(t)$

$$J(t) = J_p(t) + J_d(t). \qquad (10)$$

The magnitude of the photoconduction current at different times after illumination with fixed intensity was switched on was calculated on the basis of the measured distribution $E=E(x)$ at these times (Fig. 4a) with the help of Eqs. (7) and (8).

The computed values of the conduction current at different times after illumination is switched on and the experimentally measured relaxation of the photocurrent in the same sample are shown in Fig. 7. The curves $J=J(t)$ agree well with one another. This indicates that the change in the photocurrent with time in these structures is due to the change in the spatial distribution of the electric field in time. The contribution of the displacement current to the total photocurrent can be substantial only immediately after illumination is switched on.

*4.5. Change in the distribution of the electric field in the structure upon carrier injection into the crystal from an electrode.* We studied the effect of carrier injection into the crystal on the spatial distribution $E = E(x)$. The structure was illuminated from the anode side with a light flux of fixed intensity ($I$=35 mW/cm$^2$), giving rise to an effective redistribution of the field; carrier injection from the dark-electrode side was simulated by illuminating the structure from the cathode side with characteristic light of different intensities. We injected the carriers (electrons) of the same sign as the polarity of the dark electrode from the cathode side. Since before the illumination is switched on there is no carrier injection from the cathode (as was shown above), the change in the distribution of the electric-field strength in the crystal, when the illumination is switched on, is determined completely by the action of the injection currents.

We see from Fig. 8 that as the illumination intensity increases, the electric-field strength decreases at the dark electrode (cathode) and increases at the illuminated electrode (anode). The increase in the field strength near the anode is due to the accumulation of the charge of photoinjected electrons in the crystal near the anode at the interface with the TI layer. The formation of this charge is explained by the decrease in the electric field strength at the dark electrode (cathode) as a result of the redistribution of the electric field.

As the light intensity increases, the injection currents that appear from the dark-electrode side thus stabilize the electric-field distribution in the crystal.

## 5. CONCLUSIONS

1 The photoinduced change in the spatial distribution of the electric field in M(TI)S(TI)M structures in pure highresistivity cadmium telluride crystals was investigated: Methods and apparatus for recording this effect in structures based on electro-optic crystals were developed; the nature of the effect was described and the basic characteristics of the spatial distribution $E = E(x)$ and its kinetics under the action of illumination were determined;

a) the dependence of the magnitude of the effect on the parameters of the impurity levels in the crystal and the insulator layers in the structure was determined.

2. A new concept was proposed for photoelectric phenomena in high-resistivity structures M(TI)S(TI)M. The concept is based on the idea that the restructuring of the field [assuming $\tau(I)$= const] plays a determining role in these phenomena. This concept explains, from a unified standpoint, the diversity of observed photoelectric characteristics :

a) A method was developed for calculating the magnitude of the photocurrent and the shape of its relaxation curve from the distributions $E(x,t)$ assuming that there is no carrier injection from the contacts;

b) the current-versus-illumination characteristics of the structure and the shape of the photocurrent relaxation curve were calculated; the fact that they agree well with the experimentally measured values of the photocurrent shows that the injection properties of the photocontacts in structures based on pure crystals remain unchanged;

c) the effect of carrier injection into the crystal from the electrode side on the distribution of the field strength in the structure was investigated; it was shown that these processes stabilize the field distribution in the crystal as the illumination intensity increases;

d) the effect of the illumination on the current characteristics of the structure was studied; it was shown that the change in the photocurrent is due completely to the illumination-induced redistribution of the field in the crystal.

In summary, the proposed concept points the way for producing consistently structures with prescribed photoelectric properties. This opens up the possibility of practical applications of these phenomena for new devices and scientific methods.

[1] E. N. Arkad'eva, L. V. Maslona, O. A. Matveev. S. V. Prokof'ev, S. M. Ryvkin. and A. Kh. Khusamov, Dokl. Akad. Nauk SSSR 221. 77 (1975) [Sov. Phys. Dokl. 20. 211 (1975)].

[2] P. G. Kashenninov. A. V. Kichaev. and I. D. Yaroshelskii. Pis'ma Zh. Tekh. Fiz. 19. 48 (1993) [Tech. Phys. Leu. 19, 151 (1993)].

[3] P. G. Kashenninov. A. V. Kichaev. A. A. Tomasov. and I. D. YaroshetskiT, Pis'ma Zh. Tekh. Fiz. 20. 16 (1994) [Tech. Phys. Lett. 20. 140 (1994)].

[4] P. G. Kashenninov. D. G. Matyukhin, and I. D. Yaroshetskil, Pis'ma Zh. Tekh. Fiz. 21. 44 [Tech. Phys. Lett. 21, 260 (1995)].

[5] P. G. Kasherininov, A. V. Kichaev, and I. D. YaroshetskiT. Zh. Tekh. Fiz. 65. 193 (1995) [Sov. Phys. Tech. Phys. 40. 970 (1995)].

[6] R. H. Bube. *Photoconductivity of Solids,* Wiley, N. Y.. 1960 [Russian translation. Inostr. lit.. Moscow. 1962].

[7] A. Rose, *Concepts in Photoconductivity Theory and Allied Problems,* Wiley, N. Y., 1963 [Russian translation, Mir, Moscow, 1966].

[8] S. M. Ryvkin, *Photoelectric Effects in Solids,* Consultants Bureau, N. Y., 1964 [Russian original, Fizmatgiz, Moscow, 1963].

[9] A. G. Milnes, *Deep Impurities in Semiconductors,* Wiley, N. Y., 1973 [Russian translation, Mir, Moscow, 1976].

[10] V. I. Strikha, E. V. Buzaneva, and I. A. Radzievskaya, *Semiconductor Devices with Schottky Barriers* [in Russian], Sov. radio, Moscow, 1974.

[11] N. Mott and R. Gumey, *Electronic Processes in Ionic Crystals,* Dover Publications, N. Y., 1964 [Russian translation, Inostr. lit., Moscow, 1950].

[12] M. Lam pert and P. Mark, *Current Injection in Solids,* Academic Press, N. Y., 1970 [Russian translation, Mir, Moscow, 1973].

[13] P. G. Kasherininov, B. I. Reznikov, and G. V. Tsarenkov, Fiz. Tekh Poluprovodn. 26, 1480 (1992) [Sov. Phys. Semicond. 26, 832 (1992)]. [14] M. A. Green and J. Shewchun, Solid State Electron. 17, 349 (1974).

[15] M. A. Green, V. A. K. Temple, and J. Shewchun, Solid State Electron. 18, 745 (1975).

[16] A. A. Gutkin and V. E. Sedov, Fiz. Tekh. Potuprovodn. 9, 1761 (1975) [Sov. Phys. Semicond. 9. 1155 (1975)].

[17] A. Ya. Vul', V. I. Fedorov, Yu. F. Biryulin, Yu. S. Zinchik, S. V. Kozyrev, I. I. Saldyshev, and K. V. Sanin, Fiz. Tekh. Poluprovodn. 15, 525 (1981) [Sov. Phys. Semicond. 15, 297 (1981)].

[18] A. Ya. Vul', S. V. Kozyrev, and V. I. Fedorov, Fiz. Tekh. Poluprovodn. 15, 142 (1981) [Sov. Phys. Semicond. 15, 83 (1981)].

[19] A. Ya. Vul' and A. V. Sachenko, Fiz. Tekh. Poluprovodn. 17, 1361 (1983) [Sov. Phys. Semicond. 17, 865 (1983)].

[20] C. Cavalleri, E. Gatti, G. Fabri, and V. Svelto, Nucl. Instrum. Methods 92, 137(1971).

[21] B. S. Muravskil, V. I. Kuznetsov, G. I. Frizen. and V. N. Chemyi, Fiz. Tekh. Poluprovodn. 6. 2114 (1972) [Sov. Phys. Semicond. 6, 1797 (1972)].

[22] N. V. Agrinskaya and E. N. Arkadeva, Phys. Status Solidi B 142. K103 (1987).

[23] M. Hage-Ali, R. Stuck, C. Scharager. and P. Siffert. IEEE Trans. NS-26. 281 (1979).

[24] E. R. Mustel' and V. N. Parygin. *Methods of Modulation and Scanning of Light* [in Russian], Nauka. Moscow. 1970.

[25] A. A. Gutkin and V. E. Sedov, Fiz. Tekh. Poluprovodn. 9. 1761 (1975) [Sov. Phys. Semicond. 9, 1155 (1975)].

[26] G. Fabri and V. Svelto. Nucl. Instrum. Methods 35, 33 (1965).